\documentclass{sig-alternate-05-2015}
\usepackage{url}

\usepackage{amsmath}
\usepackage{amssymb}
\usepackage{graphicx}
\usepackage{array}
\usepackage{url}
\usepackage{color}
\usepackage{multirow}
\usepackage{booktabs}
\usepackage{verbatim}
\usepackage[english]{babel}
\usepackage{colortbl}
\usepackage[utf8]{inputenc}

\usepackage{blindtext}
\usepackage{etoolbox}
\makeatletter
\patchcmd{\maketitle}{\@copyrightspace}{}{}{}
\makeatother
\begin{document}



\clubpenalty=10000
\widowpenalty=10000

\title{Analyzing the Targets of Hate in Online Social Media\titlenote{This is a preprint of an article appearing at \textit{AAAI ICWSM 2016}.}}
%
%
%
%
%

\numberofauthors{5} 
%
\author{
%
%
\alignauthor
Leandro Silva\\
       \affaddr{UFMG, Brazil}\\
       \email{\large leandroaraujo@dcc.ufmg.br}
\alignauthor
Mainack Mondal\\
       \affaddr{MPI-SWS, Germany}\\
       \email{\large mainack@mpi-sws.org}
\alignauthor
Denzil Correa\\
       \affaddr{MPI-SWS, Germany}\\
       \email{\large denzil@mpi-sws.org}
\and  
\alignauthor
Fabr\'icio Benevenuto\\
       \affaddr{UFMG, Brazil}\\
       \email{\large fabricio@dcc.ufmg.br}
\alignauthor
Ingmar Weber\\
       \affaddr{QCRI, Qatar}\\
       \email{\large iweber@qf.org.qa}
}
\date{March 2016}

\maketitle

\begin{abstract}
Social media systems allow Internet users a congenial platform to freely express their thoughts and opinions. Although this property represents incredible and unique communication opportunities, it also brings along important challenges. Online hate speech is an archetypal example of such challenges.  Despite its magnitude and scale, there is a significant gap in understanding the nature of hate speech on social media. In this paper, we provide the first of a kind systematic large scale measurement study of the main targets of hate speech in online social media. To do that, we gather traces from two social media systems: Whisper and Twitter. We then develop and validate a methodology to identify hate speech on both these systems. Our results identify online hate speech forms and offer a broader understanding of the phenomenon, providing directions for prevention and detection approaches.
\end{abstract}

\section{Introduction}\label{sec:intro}

\noindent Social media platforms allow communication at near zero marginal cost to users. Any user with an inexpensive Internet connection has the potential to  broadcast any sort of message in these systems and reach millions of users in a short period. This property has democratized content publication: anyone can publish content, and anyone interested in the content can obtain it. This democratization has been responsible for major changes in our society. First, users can quickly gain access to information of professionals and amateurs and second, users have fewer barriers to artistic expression, benefiting from strong support for sharing creations with others.

The transformative potential of social media systems brings together many challenges.
A vivid example of such challenge is maintaining a complex balance between freedom of expression and the defense of human dignity, as these systems open  space for discourses that are harmful to certain groups of people. This challenge manifests itself with a number of variations, out of which online hate speech has
been rapidly recognized as a serious problem by the authorities of many countries. For example, the Council of Europe supports an initiative called \emph{No hate speech Movement}\footnote{\url{http://www.nohatespeechmovement.org/}}. UNESCO released a study~\cite{unecso-2015-hatespecch} entitled  \emph{Countering Online Hate Speech} aiming at helping countries to deal with the problem.

It is not surprising that most, if not all, existing efforts in this field are motivated by the impulse to ban hate speech in all forms. Existing efforts are focused on studying posts of known hate groups or radical forums~\cite{antisocial-2014-demos,burnap-2015-hatespeech,Djuric:2015:HSD:2740908.2742760,gitari2015lexicon,warnerhirschberg2012}. Only a few efforts approached this problem in systems like Twitter, but they focus on specific manifestations of the problem, like racism~\cite{twitterhate-2015-choudhry}. While these efforts are of great  importance, they do not provide the big picture about the problem in the current popular social media systems as they are usually focused on detecting specific forms of hate speech.
In this paper we take a first step towards better understanding the different forms of online hate speech. Our effort consists of characterizing hate speech in common social media, focusing on identifying the main targets of hateful messages. To do this we gathered one-year data from two social media systems: Whisper and Twitter. Then, we
propose and validate a simple yet effective method to detect hate speech using sentence structure which we used to construct our hate speech datasets.
Then, we provide the first of a kind characterization study
focused on quantitavely identifying the main targets of hate speech in these two social media systems.
Our results identify hate speech forms and unveil a set of important patterns, providing not only a broader understanding of the phenomenon, but also offering directions for prevention and detection approaches.

\section{Datasets}
\label{sec:dataset}

\noindent Next, we briefly describe how we gathered data for our study from two popular online social media sites: Whisper and Twitter.



\subsection{Data from Whisper}



Whisper is a popular anonymous social media sites, launched in March 2012 as a mobile application. Whisper users post anonymous short text  messages called ``whispers'' in this platform.  Recent work~\cite{CorreaSilva:2015,Wang-2014-whisper} suggests that users present a disinhibition complex in Whisper due to the anonymous setting. 
This property combined with its large popularity, makes Whisper an ideal environment for our study.

Whisper users can only post messages via mobile phones, however Whisper has a read only-web interface. In order to collect data from Whisper we employ a similar methodology as~\cite{Wang-2014-whisper}. We gather our dataset for one year (from 6th June, 2014 to 6th June 2015) via the ``Latest'' section of the Whisper website which contains a stream of publicly posted latest whispers. Each downloaded whisper contains the text of the whisper, location, timestamp, number of hearts (favorites), number of replies and username.


Overall, our dataset contains 48.97 million whispers posted over the year. For simplyfying further analysis we consider only the Whispers written in English and containing a valid location information. After this filtering step, we end up with \textbf{27.55 million whispers}. This corresponds to our final Whisper dataset used in the next sections.


\subsection{Data from Twitter}


\noindent Since we want to study  hate speech in the online world, along with Whisper we also collected and analyzed data from Twitter, as it is one
of the most popular social media sites today with more than 300 million monthly active users.
 The main difference between Whisper and Twitter is that users post in Twitter non-anonymously. 
In spite of the non-anonymity, there is recent evidence of hate speech in Twitter~\cite{twitterhate-2015-choudhry}. Thus, we note than it is useful to include Twitter in our study.

We collected the 1\% random sample of all publicly available Twitter data using the Twitter streaming API for a period of 1 year -- June 2014 to June 2015.  


In total we collected 1.6 billion tweets (posts in Twitter) during this period. Unlike Whisper, the default setting in Twitter is not to add location to the posts. Consequently we concentrate on only English tweets (both with and without location) posted between June 2014 to June 2015. There were \textbf{512 million tweets} in our resulting tweet dataset.

\section{Measuring Hate Speech}\label{sec:finding_hatespeech}

\noindent Before presenting our approach to measure online hate speech, we provide a few definitions.
Hate speech lies in a complex nexus with freedom of expression, group
rights, as well as concepts of dignity, liberty, and equality~\cite{unecso-2015-hatespecch}. For this reason, any objective definition (i.e., that can be easily implemented in a computer program) can be  contested. We define hate speech as \emph{any offense motivated, in whole or in a part, by the offender's bias against an aspect of a group of people}. Under this definition, online hate speech may not necessarily be a crime, but still harm people. The offended aspects can encompass basic hate crimes\footnote{https://www.fbi.gov/about-us/investigate/civilrights/\\hate\_crimes}, such as race, religion, disability, sexual orientation, ethnicity, or gender, but may also include behavioral and physical aspects that are not necessarily crimes. We do not attempt to separate organized hate speech from a rant as it is hard to infer individuals' intentions and the extent to which a message will harm an individual.

\subsection{Using sentence structure to capture hate}\label{sec:user_intent}

\noindent Most existing efforts to measure hate speech require knowing the hate words or hate targets apriori~\cite{kwokwang2013content}. Differently, we propose a simple yet very effective method for identifying hate speech in social media posts that is in agreement with our definition of hate speech and that allows us to answer our research questions.
The key idea is the following: If some user posts about their hateful emotions in a post, e.g., ``I really hate black people'', then there is little ambiguity that it is a hate post. In other words we can leverage the sentence structure to detect hate speeches with high precision very effectively. Clearly, this strategy does not identify all the existing hate speech in social media, which is fine given the purpose of the analysis presented in this study. Based on this idea, we construct the following basic expression to search in social media posts:

\begin{equation*}
I <intensity> <user intent>  <hate target>
\end{equation*}

The components of this expression are explained next.  The subject ``I'' means that the social media post matching this expression is talking about the user's personal emotions. The verb, embodied by the $<$user intent$>$ component specifies what the user's intent is, or in other word how he feels. Since we are interested in finding hate in social media posts, we set the $<$user intent$>$ component as ``hate'' or one of the synonyms of hate collected from an online dictionary\footnote{\url{http://www.thesaurus.com/browse/hate/verb}}. 
Some users might try to amplify their emotions expressed in their intent but using qualifiers (e.g., adverbs), which is captured by the $<$intensity$>$ component. Note that users might decide to not amplify their emotions and this component might be blank. Further the intensity might be negative which might disqualify the expression as a hate speech, e.g., ``I don't hate X''. To tackle this, we manually inspect the intent expressions  found using our dataset and remove the negative ones. 
Table~\ref{table:hateintent:top10} shows the top ten hate expressions formed due to the $<$intensity$>$ component in conjunction with synonyms of hate. Although the simple expression ``I hate'' accounts for the majority of the matches, we note that the use of intensifiers was responsible for 29.5\% of the matches in Twitter and for 33.6\% in Whisper. The final part in our expression is related to the hate targets.

\begin{table} [h]
	\centering
	\scalebox{0.96}{
	\begin{tabular}{|l|r|l|r|}
		\hline
        \small
		\textbf{Twitter} & \textbf{\% posts} & \textbf{Whisper} & \textbf{\% posts} \\
				\hline
				I hate & 70.5 & I hate & 66.4 \\
				\hline
				I can't stand & 7.7 & I don't like & 9.1 \\
				\hline
				I don't like & 7.2 & I can't stand & 7.4 \\
				\hline
				I really hate & 4.9 & I really hate & 3.1 \\
				\hline
				I fucking hate & 1.8 & I fucking hate & 3.0 \\
				\hline
				I'm sick of & 0.8 & I'm sick of & 1.4 \\
				\hline
				I cannot stand & 0.7 & I'm so sick of & 1.0 \\
				\hline
				I fuckin hate & 0.6 & I just hate & 0.9 \\
				\hline
				I just hate & 0.6 & I really don't like & 0.8 \\
				\hline
				I'm so sick of & 0.6 & I secretly hate & 0.7 \\
				\hline	
\end{tabular}
	}
	\caption{\textbf{Top ten hate expressions in Twitter and Whisper.}}
\label{table:hateintent:top10}
\end{table}

\noindent \textbf{Determining hate targets:} A simple strategy that searches for the sentence structure $I$ $<$intensity$>$ $<$user intent$>$ $<$any word$>$ results in a number of posts that do not contain hate messages against people, i.e., ``I really hate owing people favors'', which is not in agreement with our the definition of online hate speech considered in our work. Thus, to focus on finding hate against groups of people in our datasets, we design two templates for the hate target component.

We design the first template for our $<$hate target$>$ token as simply \textit{``$<$one word$>$ people''}. For example we search for patterns like ``black people'' or ``mexican people''. This template for $<$hate target$>$ captures when hate is directed towards a group of people. However we observe that even with this template we found some false positives as there are posts like ``I hate following people''. To reduce the number of such false positives we create a list of exclusion words for this approach including words like following, all, any, watching, when, etc.

Second, not all hate words come together with the term ``people''. To account for this general nature of hate speech we employ the help of Hatebase~\footnote{\url{http://www.hatebase.org/}}. It is the world's largest online repository of structured, multilingual, usage-based hate speech. Hatebase uses crowdsourcing to build its collection of hate words. We crawled Hatebase on September 12, 2015 to collect a comprehensive list of hate words. There are 1,078 hate words in Hatebase spanning eight categories: archaic, class, disability, ethnicity, gender, nationality, religion, and sexual orientation. However each word  in Hatebase comes with an offensivity score. The score varies from 0 to 100, with 100 indicating most offensive hate words. Since our goal is to find serious hate speech from social media data we take only the hate words from Hatebase with offensivity greater than 50\%\footnote{There are 116 such hate words in Hatebase}, and use those words as template for $<$hate target$>$ tokens in our pattern.

\begin{table}[htb!]
\centering\small
\begin{center}
\begin{tabular}{|p{2.5cm}|r|p{2cm}|r|}
		\hline
		\multicolumn{2}{|c|}{\emph{\textbf{Twitter}}} & \multicolumn{2}{|c|}{\emph{\textbf{Whisper}}} \\
				\hline
				{\bf Hate target} & {\bf \% posts} & {\bf Hate target} & {\bf \% posts} \\
				\hline
				\hline
				Nigga &  31.11 & Black people & 10.10 \\
				\hline
				White people & 9.76 & Fake people & 9.77 \\
				\hline
				Fake people & 5.07 & Fat people & 8.46 \\
				\hline
				Black people & 4.91 & Stupid people & 7.84 \\
				\hline
				Stupid people & 2.62 & Gay people & 7.06 \\
				\hline
				Rude people & 2.60 & White people & 5.62 \\
				\hline
				Negative people & 2.53 & Racist people & 3.35 \\
				\hline
				Ignorant people & 2.13 & Ignorant people & 3.10 \\
				\hline
				Nigger & 1.84 & Rude people & 2.45 \\
				\hline
				Ungrateful people & 1.80 & Old people & 2.18 \\
				\hline
\end{tabular}
\end{center}
\caption{\textbf{ Top ten targets of hate in Twitter and Whisper}}
\label{table:targets_across_nets}
\end{table}

Overall, our strategy was able to identify \textbf{20,305 tweets} and \textbf{7,604 whispers} containing hate speech. We present the top hated targets from Twitter and Whisper using our methodology in Table~\ref{table:targets_across_nets}. It shows that, racist hate words like ``Black people'', ``White people'' or ``Nigga'' are the most used hate targets. We further checked how many of these hate messages are detected by our two different templates. 
Overall, the template with ``people'' finds more hate speech than using the words from Hatebase, accounting for 65\% of the Twitter dataset and 99\% of the Whisper dataset.
One possible reason for this high difference in the two datasets is that Whisper operators are already filtering out posts containing some of the words from Hatebase.


\subsection{Evaluating our hate speech detection methodology}\label{sec:evaluation:methodology}

\noindent Next, we evaluate the precision of our hate speech detection approach. To that end we did a simple experiment: We randomly sample 100 posts of all the whispers which matched our language structure based expression. Then we manually verify whether these 100 posts are really classified as hate speech by human judgment. We observe that 100\% of both the whispers and tweets can be classified as hate speech by human judgment, where the poster expressed their hate against somebody. It is important to highlight that our methodology was not designed to capture \textit{all} or most of the hate speech that in social media. In fact, detecting online hate speech is still an open research problem. Our approach aims at building a dataset that allow us to identify the main targets of online hate speech.



\subsection{Categorizing Hate Speech}\label{sec:categorizing}

\noindent Our final methodological step consists of manually categorizing hate targets. For example, the term ``black'' should be categorized as race and ``gay'' as sexual orientation. In order to decide the hate categories we take inspiration from Hatebase. Hatebase along with the words gave us hate categories like ethnicity, race, religion, etc. We also consider categories reported by FBI for hate crimes. We combine these two sets of categories and added two more for better coverage of our data. We end up with nine categories: Race, Behavior, Physical, Sexual orientation, Class, Gender,  Ethnicity, Disability, and Religion. We also add an ``other'' category for any non-classified hate targets. The final hate categories and some examples of hate targets for each category are in Table~\ref{table:hatecat:example}.

\begin{table}[htb!]
\centering\small
\begin{center}
\begin{tabular}{|p{3cm}|p{4.5cm}|}
\hline
{\bf  Categories} & {\bf Example of hate targets}\\
\hline
\hline
 Race  &  nigga, black people, white people \\
 \hline
 Behavior  &  insecure people, sensitive people \\
 \hline
 Physical  &  obese people, beautiful people\\
 \hline
 Sexual orientation  &  gay people,  straight people\\
 \hline
 Class  &  ghetto people, rich people\\
 \hline
 Gender  &  pregnant people, cunt, sexist people \\
 \hline
 Ethnicity  &  chinese people, indian people, paki\\
 \hline
 Disability  &  retard, bipolar people \\
 \hline
 Religion  &  religious people, jewish people \\
 \hline
 Other  &  drunk people, shallow people \\
\hline

\end{tabular}
\end{center}
\caption{{\bf Hate categories and example of hate targets.} }
\label{table:hatecat:example}
\end{table}

Since manual classification of hate targets are resource consuming, we aim to categorize only the top hate words that cover most of the hate speech in our data. In total, the Twitter and the Whisper datasets contain 264 and 242 unique hate targets, respectively, and most of them appear in both datasets.
We  manually label the most popular 178 targets, which account for 97\% of the Twitter data and also 97\% of the whispers.  In the next section we look into these hate categories and the associated hate speech from them more in depth.

\section{Targets of Online Hate Speech}

\noindent We start with observing which categories of hate are most prevalent in our experimental platforms -- Twitter and Whisper. The results are shown in Table~\ref{table:categories_across_nets}. The hate categories are sorted by the number of hate speech in these categories (except for the non-classified hate targets, which we put in the other category). We made two interesting observations from this table. First, for both Twitter and Whisper the top three hate categories are the same -- Race, behavior, and physical. However, in Twitter these categories cover 89\% of the tweets, whereas in Whisper they cover only 69\% of all the whispers related to hate. One potential explanation for this difference may be due some existing filtering that Whisper might already apply for very aggressive hate words, like those from Hatebase. We also note that for these categories in both Twitter and Whisper, there is also hate as a response to hate, e.g., ``I hate racist people''. However such types of hate are not expressed in a high number of posts, and hate with negative connotation is more common.

Secondly we observe that out of the top three hate categories for both Twitter and Whisper, the categories behavior and physical aspects are more about \textit{soft} hate targets,
like fat people or stupid people. This observation suggests  that perhaps a lot of online hate speech is targeted towards groups of people that are not generally included when documenting offline hate crimes. For e.g., https://www.fbi.gov/news/stories/2015/november/latest-hate-crime-statistics-available/ contains a breakdown of offline hate crimes according to their bias. 

\begin{table}[htb!]
\centering\small
\begin{center}
\scalebox{0.95}{
\begin{tabular}{|p{2.3cm}|p{1.15cm}|p{2.3cm}|p{1.15cm}|}
		\hline
		\multicolumn{2}{|c|}{\emph{\textbf{Twitter}}} & \multicolumn{2}{c|}{\emph{\textbf{Whisper}}} \\
				\hline
				{\bf Categories} & {\bf \% posts} & {\bf Categories} & {\bf \% posts} \\
				\hline
				\hline
				Race & 48.73 & Behavior & 35.81 \\
				\hline
				Behavior & 37.05 & Race & 19.27 \\
				\hline
				Physical & 3.38 & Physical & 14.06 \\
				\hline
				Sexual orientation & 1.86 & Sexual orientation & 9.32 \\
				\hline
				Class & 1.08 & Class & 3.63 \\
				\hline
				Ethnicity & 0.57 & Ethnicity & 1.96 \\
				\hline
				Gender & 0.56 & Religion & 1.89 \\
				\hline
				Disability & 0.19 & Gender & 0.82 \\
				\hline
				Religion & 0.07 & Disability & 0.41 \\
				\hline
				Other & 6.50 & Other & 12.84 \\
				\hline				
\end{tabular}
}
\end{center}
\caption{\textbf{Hate categories distribution.}}
\label{table:categories_across_nets}
\end{table}

\section{Conclusion}
\label{sec:conclusion}

\noindent The fight against perceived online hate speech is beginning to reach a number of concerned parties, from governments to private companies,
as well as to a growing number of active organizations and affected individuals. Our measurement study about online hate speech provides an overview of how this very important  problem of the modern society currently manifests.  Our effort even unveils new forms of online hate that are not necessarily crimes, but can be harmful to people. We hope that our dataset and methodology can help  monitoring systems and detection algorithms to identify novel keywords related to hate speech as well as inspire more elaborated mechanisms to identify online hate speech.  Building a hate speech detection system that leverages our findings is also part of our future research agenda.

\section{Acknowledgments}
This work was partially supported by the project FAPEMIG-PRONEX-MASWeb,
Models, Algorithms and Systems for the Web, process number APQ-01400-14, and grants from CNPq, CAPES, and Fapemig.

\bibliographystyle{abbrv}
\bibliography{silva-mondal}

%
%
\end{document}